\documentclass[twocolumn,aps,showpacs,prl,superscriptaddress,notitlepage,longbibliography]{revtex4}
\usepackage[colorlinks=true,urlcolor=blue,citecolor=blue,linkcolor=blue]{hyperref}
\usepackage{graphicx}  
\usepackage{dcolumn}   
\usepackage{bm}        
\usepackage{amssymb}
\usepackage{amsfonts}
\usepackage{doi}
\usepackage{verbatim}
\usepackage{epstopdf}

\usepackage{amsmath}
\usepackage{braket}
\usepackage{hyperref}

\DeclareMathOperator{\diag}{diag}
\begin{document}

\title{Evolution of Berry curvature and Reentrant Quantum Anomalous Hall Effect in an Intrinsic Magnetic Topological Insulator}
\author{Chui-Zhen Chen}
\affiliation{Institute for Advanced Study and School of Physical Science and Technology, Soochow University, Suzhou 215006, China.}
\author{Junjie Qi}
\affiliation{Beijing Academy of Quantum Information Sciences, West Bld.3,
No.10 Xibeiwang East Rd., Haidian District, Beijing 100193, China}
\author{Dong-Hui Xu}
\affiliation{Department of Physics, Hubei University, Wuhan 430062, China. }
\author{X. C. Xie}\thanks{xcxie@pku.edu.cn}
\affiliation{International Center for Quantum Materials, School of Physics, Peking University, Beijing 100871, China}
\affiliation{CAS Center for Excellence in Topological Quantum Computation,
University of Chinese Academy of Sciences, Beijing 100190, China}
\affiliation{Beijing Academy of Quantum Information Sciences, West Bld.3,
No.10 Xibeiwang East Rd., Haidian District, Beijing 100193, China}
\begin{abstract}
Recently, the magnetic topological insulator (TI) MnBi$_2$Te$_4$ emerged as a competitive
platform to realize quantum anomalous Hall (QAH) states. We report a Berry-curvature splitting mechanism to realize the QAH effect in the disordered magnetic TI multilayers when switching from an antiferromagnetic order to a ferromagnetic order. We reveal that the splitting of spin-resolved Berry curvature, originating from the separation of the critical points during the magnetic switching, can give rise to a QAH insulator. We present a global phase diagram, and also provide a phenomenological picture to elucidate the Berry curvature splitting mechanism by the evolution of topological charges. At last, we predict that the Berry curvature splitting mechanism will lead to a reentrant QAH effect, which can be detected by tuning the gate voltage. Our theory will be instructive for the studies of the QAH effect in MnBi$_2$Te$_4$ in future experiments.\vspace{0.25cm}\\
{\bf{quantum anomalous Hall effect, Berry curvature, disorder, localization}\rm}
\end{abstract}


\maketitle

{\emph{Introduction.}}---
The quantum anomalous Hall (QAH) state was proposed as a novel quantum state that exhibits quantized Hall conductance from topologically nontrivial bands rather than Landau levels \cite{Haldane1988}. Experimentally, the QAH state was initially realized in Cr-doped (Bi,Sb)$_2$Te$_3$ thin films, where the band gap is inverted by the Zeeman splitting \cite{LiuQAH2008,Yurui2010,Chang2013,Checkelsky2014,Kou2014,Liu2016,Heke2018}. However, the QAH effect in a magnetically doped topological insulator (TI) occurs at very low temperatures due to inhomogeneity introduced by magnetic dopants.
Recently, a breakthrough was made in realization of an intrinsic magnetic TI in a van der Waals layered material MnBi$_{2}$Te$_{4}$  \cite{Heke2019,Wangjing,Lieaaw5685,otrokov2019,Vidal2018,
Hu2020,wu2019,li2019dirac,chen2019topological,vidal2019topological,Shi_2019,ding2020,
huyong2020,xu20202persistent,tian2020magnetic,hao2019gapless,wangjian2019,Yayu2020,yuanbo2020}. MnBi$_{2}$Te$_{4}$ exhibits antiferromagnetic (AFM) order and supports an axion insulator state without an external magnetic field.
Applying the magnetic field to drive a transition from an AFM order to a ferromagnetic (FM) order can result in a quantum  phase transition from an axion insulator to a QAH state in the even-number septuple-layered MnBi$_{2}$Te$_{4}$ \cite{Yayu2020,yuanbo2020,OtrokovPRL2019,Deng_2020}.
Meanwhile, the QAH state was also observed in the odd-number septuple-layered MnBi$_{2}$Te$_{4}$, but the quantization of Hall resistance  strongly depends on disorder and external magnetic fields \cite{yuanbo2020}.
This implies that disorder and magnetization are two key ingredients to determine and manipulate the QAH effect in this intrinsic magnetic TI material \cite{Anderson1958,Anderson1979,Belitz1994,Onoda2003,Qiao2016,CZChen2020,liyuhang2021} .
More importantly, it remains an open question that the onset temperature of the QAH in  MnBi$_{2}$Te$_{4}$ is still an order of magnitude smaller than the inverted band gap \cite{yuanbo2020}.
Therefore, it is crucial to understand the origin of the QAH state and disorder effects in the few-layer MnBi$_{2}$Te$_{4}$ during the magnetic switching.

In this article, we propose a mechanism of the QAH effect in the few-layer magnetic TI MnBi$_2$Te$_4$.
We reveal that the splitting of spin-resolved Berry curvature during the magnetic reversal can give rise to a QAH insulator phase even {\it without} closing the global energy gap in the disordered magnetic TI. This is in contrast to the band inversion mechanism of the clean QAH insulator, which always involves the closing and reopening of the bulk gap \cite{Yurui2010}.
We further show that the Berry curvature can only be carried by the conducting states near the critical points and thus they are separated with increasing Zeeman splitting. 
This suggests that the disorder-induced critical points play a key role in realizing the QAH state in disordered magnetic TIs. We provide a phenomenological theory based on the evolution of topological charges at the critical points to interpret the creation of the QAH phase.
At last, we predict a reentrant QAH effect, i.e. a QAH-normal insulator-QAH transition, to verify our theory in future transport experiments.
\begin{figure}[bht]
\centering
\includegraphics[width=3.2in]{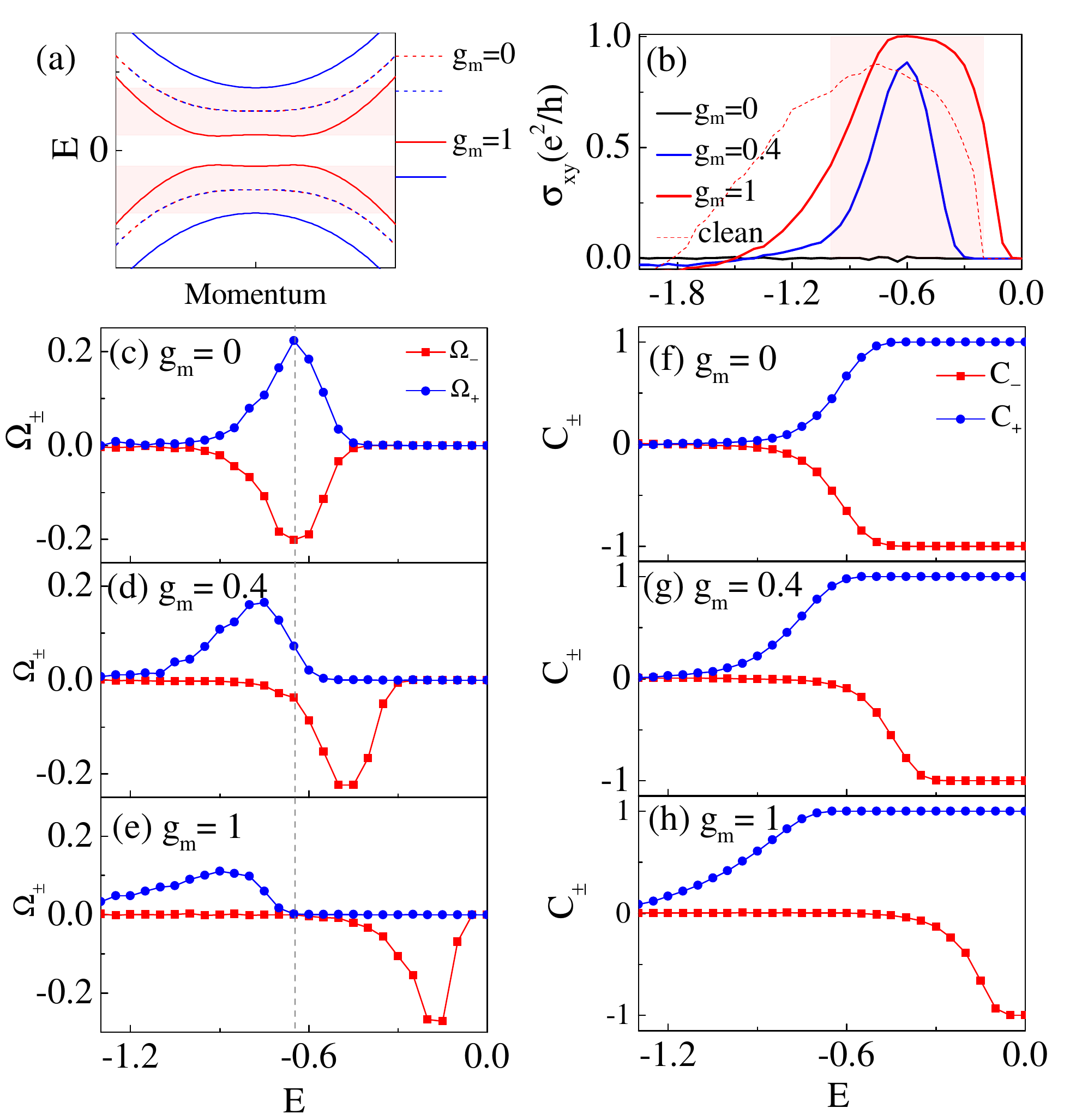}
\caption{(Color online). (a) Schematic plot of energy spectrum of the clean few-layer MnBi$_2$Te$_4$ for $g_m=0$ and $g_m=1$. (b) Hall conductance $\sigma_{xy}$ versus Fermi energy $E$ for different magnetization $g_m$. Note that the dashed red line shows $\sigma_{xy}$ of a clean FM TI sample with $g_m=1$ and $W=0$.
The red shadow region represents the disordered QAH insulator phase due to Berry curvature $\Omega_{\pm}$ splitting at $g_m=1$.
(c)-(e) The splitting of Berry curvature $\Omega_{\pm}$ and (f)-(h) the evolution of spin polarized Chern number $C_{\pm}$ versus $E$ with increasing $g_m$. All the data are averaged over 200 disorder configurations with sample size $48\times 48$.
\label{fig1} }
\end{figure}

\begin{figure}[tbh]
	\centering
	\includegraphics[width=3.2in]{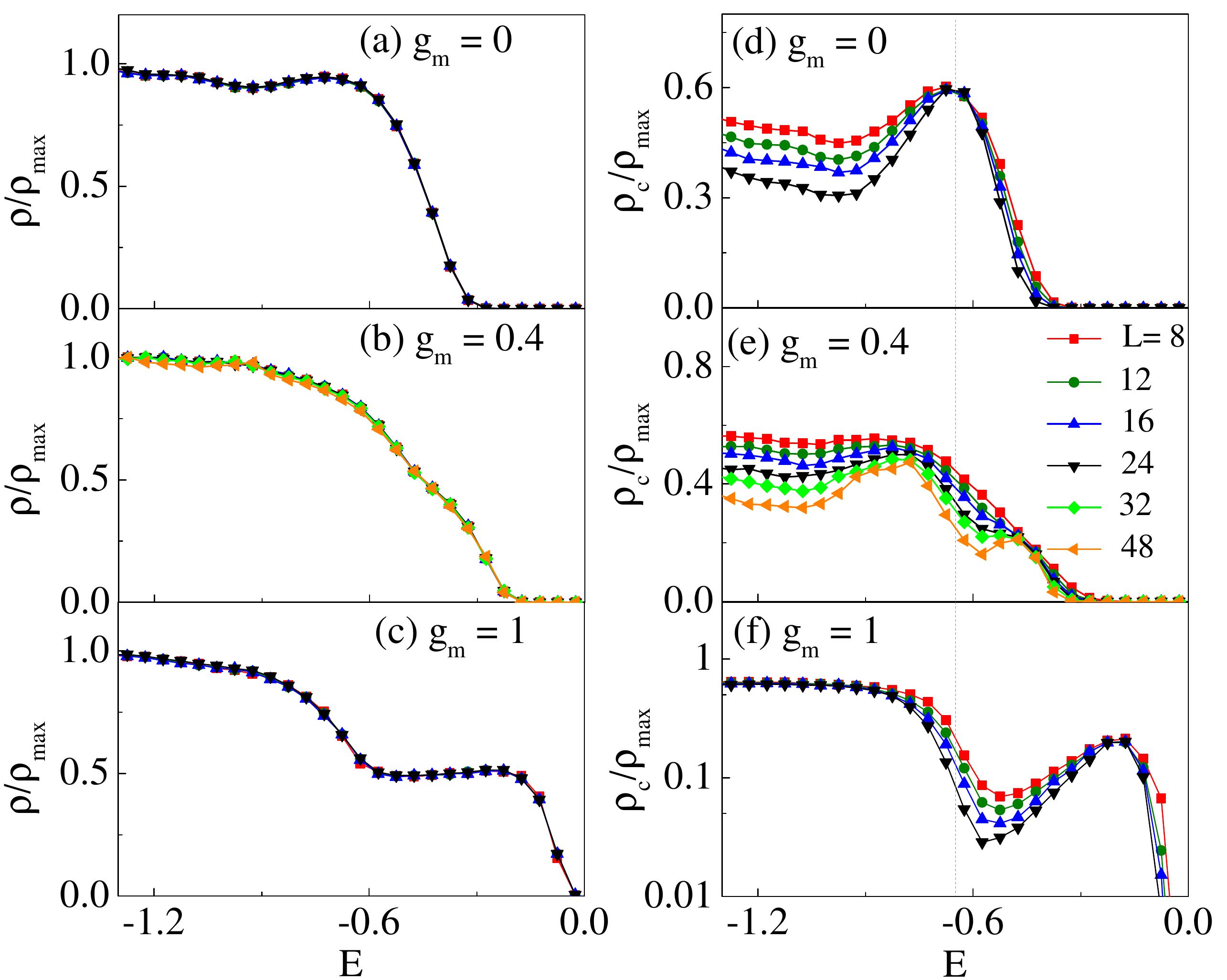}
	\caption{(Color online). (a)-(c) The total density of states $\rho$  and (d)-(e) the density of conducting states $\rho_c$  versus Fermi energy $E$ with increasing magnetization $g_m$. $\rho_{max}$ is maximum total density of states. Different line colors represent different sample sizes $L\times L$.  \label{fig2} }
\end{figure}

{\em Model Hamiltonian.}---
The $\mathbf{k} \cdot \mathbf{p}$ Hamiltonian of few-layer MnBi$_2$Te$_4$ at the $\Gamma$ point can be written as \cite{zhang2019mbius,lian2019flat,Wangjing,CZChen2021}
\begin{eqnarray}
  H &=& H_N(\mathbf{k}) + g_m  \delta H_{M}(\mathbf{k}) \nonumber
\end{eqnarray} with the nonmagnetic part $H_N(\mathbf{k})$ and the FM part $g_m\delta H_M(\mathbf{k})$ of the effective Hamiltonian given by
\begin{eqnarray*}
   H_N & = &  \epsilon_0(\mathbf{k})+ \left(
                        \begin{array}{cccc}
                          M(k) &  V     & 0    & Ak_- \\
                          V    &  -M(k)  & Ak_- & 0      \\
                          0    &  Ak_+  & M(k)& V \\
                          Ak_+ &  0     & V    & -M(k) \\
                        \end{array}
                      \right)
\end{eqnarray*} and
\begin{eqnarray*}
    \delta H_{M} & = & \left(
                        \begin{array}{cccc}
                          M_1(k) &  0       & 0       & A'k_- \\
                          0      &  M_2(k)  & -A'k_- & 0      \\
                          0      &  -A'k_+ & -M_1(k) & 0 \\
                          A'k_+ &  0       & 0       & -M_2(k) \\
                        \end{array}
                      \right)
\end{eqnarray*}
in the four-orbital basis of
$|P1_z^+,\uparrow\rangle$, $|P2_z^-,\uparrow\rangle$, $|P1_z^+,\downarrow\rangle$
and $|P2_z^-,\downarrow\rangle$.
Here we introduce a dimensionless magnetization strength $0\leq g_m\leq1$ to describe the evolution from an AFM phase at $g_m=0$ to a FM phase at $g_m=1$.
The mass terms are $M(k)=m_0-B_0k^2$ and $M_{1,2}(k)=m_{1,2}-B_{1,2}k^2$,
$\epsilon_0(k)=Dk^2$ and the wave vectors are $k_\pm=k_x\pm ik_y$, with the model parameters $A$, $A'$, $B_{0,1,2}$, $D$ and $m_{0,1,2}$ \cite{lian2019flat}.
$V$ denotes inversion-symmetry breaking potential strength.
If $V=0$, the system $H_N$ reduces to the Bernevig-Hughes-Zhang model describing a 2D TI with a pair of helical edge modes when $m_0B_0>0$ \cite{Andrei2006}. Such a pair of helical edge modes has been recently observed by transport measurement in MnBi$_2$Te$_4$ multilayers \cite{ToBeSubmitted2021,Liu2020helical}, and we focus on $B_0=0.5$ and  $m_0=0.5$ in the following.
Other parameters are set as  $m_1=m_2=0.3$, $A=1.1$, $A'=0.6$, $B_{1,2}=D=0$ and $V=0.05$ unless specified  \cite{Wangjing} and $H$ is discretized into a square lattice with a lattice constant $a=1$.

Figure~\ref{fig1}(a) plots the energy spectrum of the Hamiltonian $H$ with $V=0$.
For the AFM phase with $g_m=0$, the spectrum is double degenerated (see the dashed lines) due to the coexistence of an effective time-reversal symmetry and inversion symmetry. When turning on the FM term $g_m\delta H_{M}$ with $g_m=1$,
 the two degenerate bands split and leave an energy spacing between them (see the solid blue, red lines and red shaded region).
In the following, we will show that the interplay of the band splitting and the disorder induced localization effect can give rise to a QAH insulator phase in the red shaded region without closing the global energy  gap.
Here, the magnetic disorder that dominates in magnetic TI is included as $H_D=V(r)\sigma_z$, where the random potential $V(r)\in[-W/2,W/2]$ with disorder strength $W$ and the Pauli matrix $\sigma_z$  acting on the spin degrees of freedom.

{\em QAH insulator from Berry curvature splitting.}---
To characterize the QAH effect of MnBi$_2$Te$_4$ in the presence of the magnetic disorder $H_D$,
we adopt the zero-temperature Hall conductance defined as $\sigma_{xy}=(C_+ + C_-)e^2/h$.
Here the spin-polarized Chern number $C_\pm$ is evaluated by the real-space noncommutative Kubo formula~\cite{Prodan2009,Prodan2011,Sheng2005,Sheng2006}
\begin{eqnarray}
  C_{\pm} &=& \int_{-\infty}^E  \Omega_{\pm}(\varepsilon) d\varepsilon, \\
  \Omega_{\pm}(\varepsilon) &=& 2\pi i \left\langle\operatorname{Tr}\left\{P^{\pm}_\varepsilon\left[-i\left[\hat{x}, P^{\pm}_\varepsilon\right],-i\left[\hat{y}, P^{\pm}_\varepsilon\right]\right]\right\}\right\rangle, \nonumber\label{eq:chern}
\end{eqnarray}
imposing the periodic boundary condition in both $x$ and $y$ directions. $\langle ... \rangle$ is ensemble averaged over random configurations, and ($\hat{x}$, $\hat{y}$) denotes the position operator.
$P^{\pm}_\varepsilon$ is the spectral projector onto the positive ($+$) or negative ($-$) eigenvalue of $P_{\varepsilon}^{ } \Gamma_z P_{\varepsilon}^{ }  $, where $P_\varepsilon^{ } $ is the projector onto the state $|\varepsilon\rangle$ of eigenenergy $\varepsilon$ and $\Gamma_z=\diag\{1,-1,-1,1\}$.

Figure~\ref{fig1}(b) depicts the disorder-averaged zero-temperature Hall conductance $\sigma_{xy}$ as a function of the Fermi energy $E $ for various magnetization $g_m$. We fix the disorder strength $W=3$ and only show the $E<0$ part since the $E>0$ part is the same due to particle-hole symmetry.
For the AFM phase with $g_m=0$, the system preserves an averaged time-reversal symmetry and thus the Hall conductance $\sigma_{xy}$ approaches zero. Note that the averaged time-reversal symmetry means that time-reversal symmetry, which is violated by magnetic disorder, remains unbroken on average \cite{Fu2012}.
When we turn on the magnetization $g_m$, $\sigma_{xy}$ grows and exhibits a peak near $E=0.6$ when $g_m=0.4$,
and it is fully quantized to be $e^2/h$ when $g_m=1$, meaning that a QAH insulator phase shows up (in the red region). For comparison, we also show $\sigma_{xy}$ at $g_m=1$ for a clean sample ($W=0$) (see the red dashed line). Apparently, $\sigma_{xy}$ is non-quantized in this case and the system behaviors as as an anomalous Hall metal \cite{Chen2020}.
This suggests the disorder plays an important role and can lead to the QAH insulator from a metal in the red region [see Fig~\ref{fig1}(b)].

To elucidate the origin of the QAH insulator phase in Fig.~\ref{fig1}(b), we investigate the spin-polarized Berry curvature $\Omega_{\pm}(E)$
and the spin-polarized Chern number $C_{\pm}$ in Figs.~\ref{fig1}(c-h).
At $g_m=0$, one can see that the Berry curvature $\Omega_{\pm}(E)\equiv dC_{\pm}/dE$ shows two peaks of opposite values at $E_c\approx0.65$ in Fig.~\ref{fig1}(c). Meanwhile, the spin-polarized Chern numbers $C_{\pm}$ show a Heaviside function behavior $\Theta(E-E_c)$ with $E_c\approx0.65$ in Fig.~\ref{fig1}(f).
Here the spin-polarized Chern numbers $C_{\pm}=\pm1$ originate from the topological nature of the 2D TI, which has been recently observed by transport measurement of MnBi$_2$Te$_4$ multilayers \cite{ToBeSubmitted2021,Liu2020helical}.
By increasing $g_m$, the two peaks of the Berry curvature $\Omega_{\pm}(E)$ separate as shown in Figs. \ref{fig1}(d-e)
and therefore the two Chern numbers $C_{\pm}$ with opposite sign are no longer symmetric with respect to zero [see Figs. \ref{fig1}(g-h)].
This gives rise to a net Hall conductance plateau $\sigma_{xy}=(C_- +C_+)e^2/h$ near $E\approx 0.6$ when $g_m=1$
in Fig.~\ref{fig1}(b).
As a result, we conclude that that the QAH insulator found above results from the Berry curvature splitting  caused by disorder and Zeeman splitting.

Next, let us explore the relation between the Berry curvature and the conducting states. The localized and extended states can be distinguished by smooth changes in boundary conditions of the wave function \cite{Arovas1988,Bhatt1992,Zhuqiong2020}.
The total density of states $\rho$ and the density of conducting states $\rho_c$ for a sample with size $S=L\times L$ are defined as $\rho =dN(E)/dS$ and $\rho_c=dN_c(E)/dS$, where $N(E)$ and $N_c(E)$  are the total number of states and the number of conducting states at $E$, respectively.

In Figs.~\ref{fig2}(a-c), the density of state $\rho$ spreads towards the band gap with increasing $g_m$ thanks to the splitting of two generate bands. However, the band gap does not close as the magnetization increases from $g_m=0$ to $g_m=1$, implying the QAH insulator in Fig.\ref{fig1} cannot be attributed to the band inversion.
For comparison, we plot the density of conducting states $\rho_c$ against $E$ for different $g_m$ in Figs.~\ref{fig2}(d-f).
Generally, the decrease in $\rho_c$ with the system size $L$ indicates an insulating phase since the density of conducting states goes to zero in the large $L$ limit, while $\rho_c$ does not change with $L$ at the critical point.
When $g_m=0$, there are two double generate critical points at $E=0.65$ due to the averaged time-reversal symmetry.
They split and move to $E_{c1}=0.45$ and $E_{c2}=0.75$ when $g_m=0.4$, and then to $E_{c1}=1$ and $E_{c2}=0.2$ when $g_m=1$. This leads to a mobility gap between two critical points since $\rho$ is nonzero [see Figs.~\ref{fig2}(b-c)]. The quantized Hall conductance region in Fig.~\ref{fig1}(b) is exactly the same as the mobility gap region, which verifies that the QAH insulator results from the splitting the Berry curvature  {\it without} closing the global energy gap in the disordered magnetic TI.
We would like to emphasize that the critical point positions of $\rho_c$ are consistent with the peak positions of the Berry curvatures $\Omega_{\pm}$ in Figs.~\ref{fig1}(c-e). This demonstrates that the Berry curvature $\Omega_{\pm}$ is only carried by the conducting states.


\begin{figure}[tbh]
\centering
\includegraphics[width=3.4in]{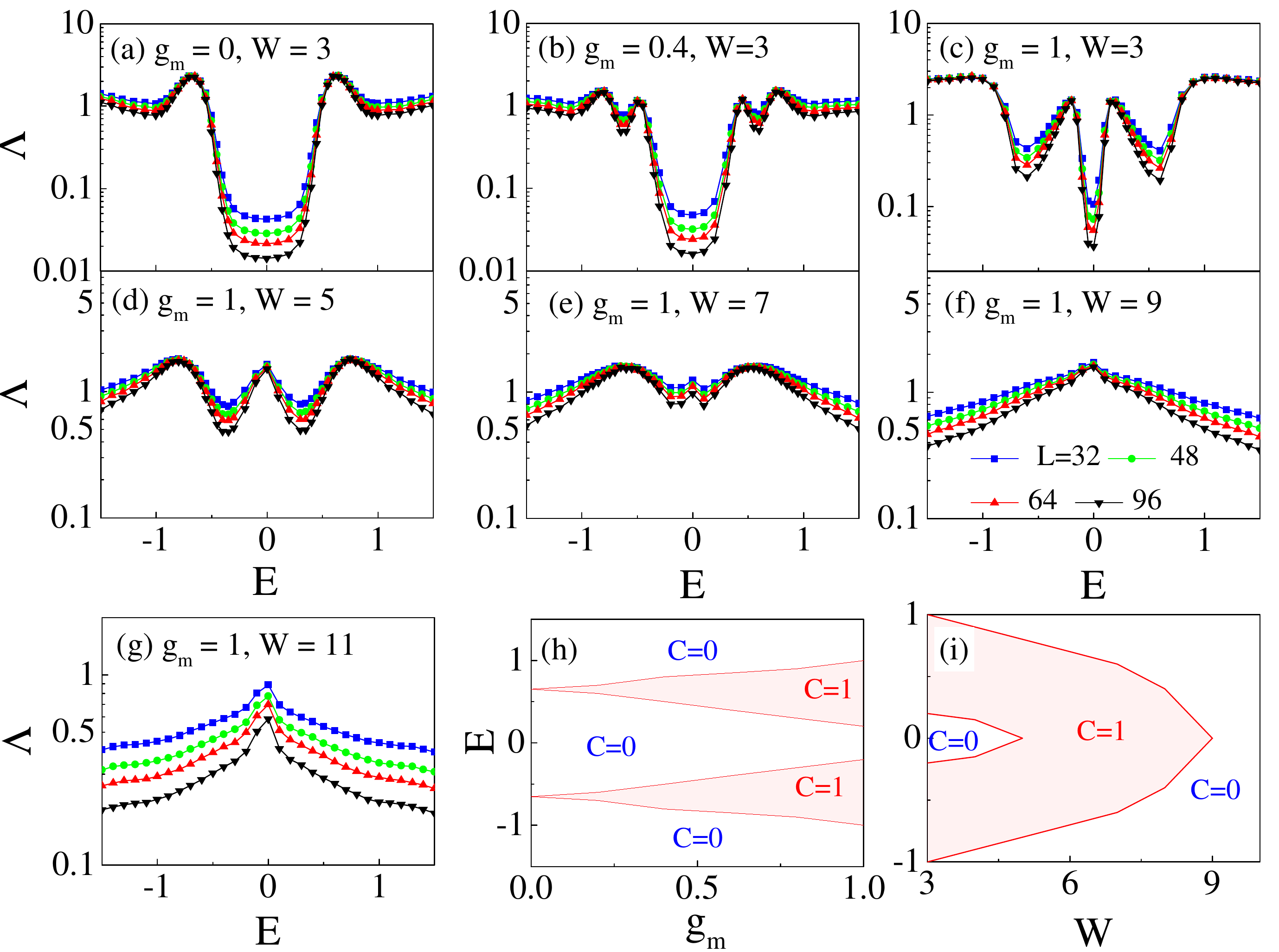}
\caption{(Color online). (a)-(g) Normalized localization length $\Lambda=\lambda/L$ versus energy $E$
at different magnetization $g_m$ and disorder strength $W$.
The curves correspond to different sample widths $L$.
The phase diagrams of disorder magnetic TI on the (h)$g_m$-$E$  plane with $W=3$ and (i)$W$-$E$ plane with $g_m=1$.
In (h), one can see that two QAHI phase with $C=1$ (red regions) show up with increasing $g_m$ due to the Berry curvature splitting, while the global band gap does not close. $D=0.01$ and other parameters are the same as Fig.~\ref{fig2}.
\label{fig3} }
\end{figure}
{\em Phase diagram and phenomenological theory.}--
To gain the global phase structure, we perform the finite-size
scaling analysis of the localization length to obtain the phase diagrams of the magnetic TI.
The localization length $\lambda$ can be evaluated by the standard transfer matrix method using a long cylinder sample of $L \times L_y$  with the periodic boundary condition in the $x$ direction and the open boundary condition in the $y$ direction \cite{MacKinnon1981,MacKinnon1983,Kramer1993}. The sample length is chosen to be $L_y=10^6$.
The normalized localization length $\Lambda=\lambda/L$ increases (decreases) with the increase in width $L$ of the sample for a metallic (insulator) phase, and $d\Lambda/dL=0$ for a critical point \cite{Kramer1993}.

Figures~\ref{fig3}(a)-(c) plot the normalized localization length $\Lambda$ against the energy $E$ for different magnetization strengths at $W=3$. The conduction and valence bands both possess two degenerate critical points
with $d\Lambda/dL=0$ at $g_m=0$ and they separate apart with increasing $g_m$. This is in accordance with the scaling behaviors of the density of conducting states $\rho_c$.
When we further increases disorder strength $W$ for $g_m=1$,
the critical points move toward the band center [see Figs.~\ref{fig3}(d-g)].
One pair of the critical points annihilate at the band center when $W=5$ [see Fig.~\ref{fig3}(d)], while the other pair survive
for much stronger disorder and eliminate until $W=9$ [see Fig.~\ref{fig3}(f)].
We summarize the main results in the phase diagrams in Figs.~\ref{fig3}(h)-(i).
It is clear that a pair of QAH phases with $C=1$ in the conduction and valence bands are created, and the energy window of the QAH phases is expanded with increasing $g_m$ without closing the global energy gap [see Fig.~\ref{fig3}(h)].
Then by increasing disorder strength $W$, the two QAH phases move to the band center, merge together
and at last annihilate in strong disorder limit [see Fig.~\ref{fig3}(i)].
Such a disorder-induced floating of critical points behavior was also previously discovered in quantum Hall systems \cite{DZLiu1996,XCXie1996}.
\begin{figure}[tbh]
	\centering
	\includegraphics[width=3.2in]{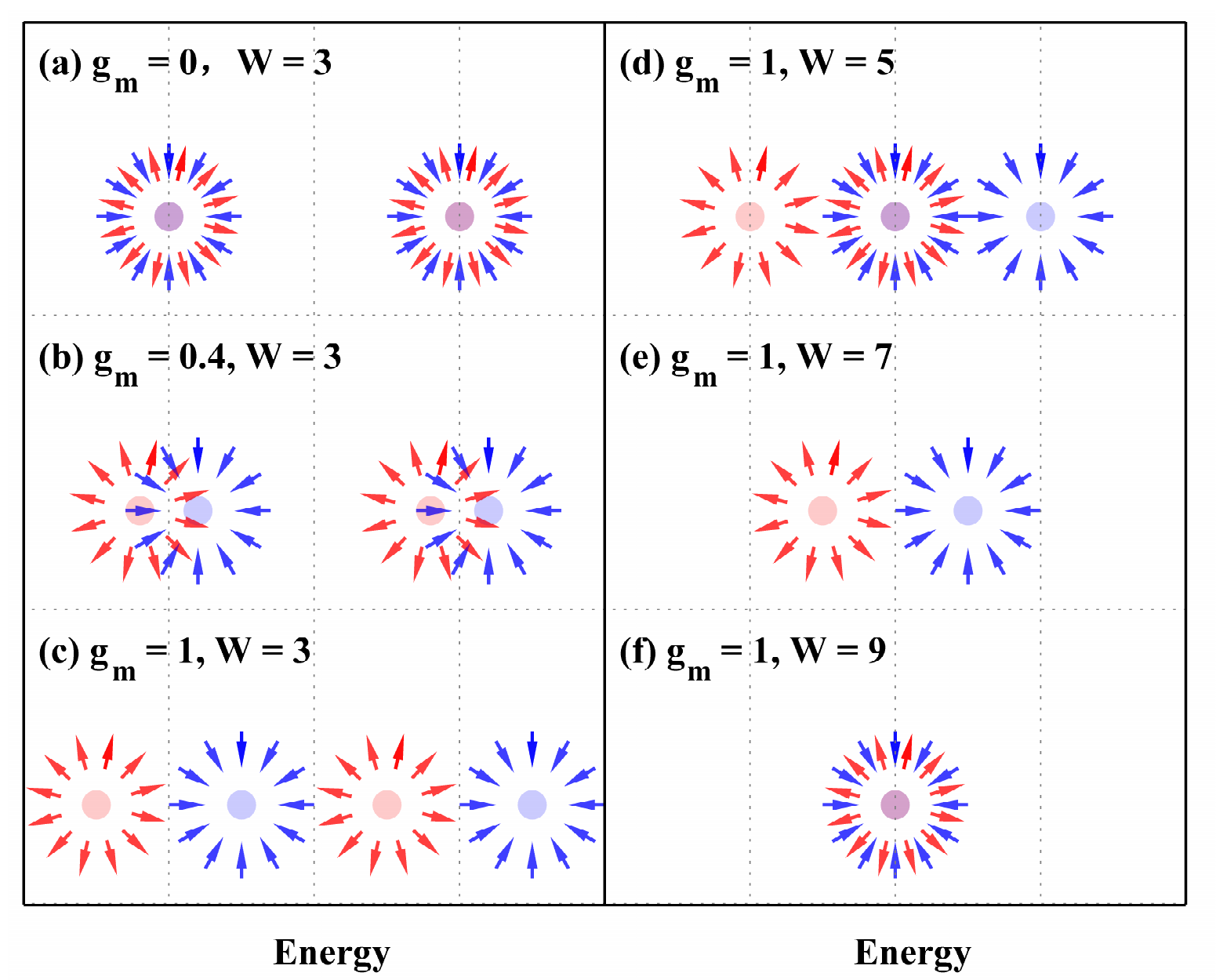}
	\caption{(Color online). Schematic plot of the skyrmions of opposite charges $Q^{v}_{1,2}=\pm 1$ ($Q^{c}_{1,2}=\mp 1$) in valence (conduction) band as spins pointing outwards and inwards the spheres.
(a) In the absence of magnetization $g_m=0$, the skyrmions and antiskyrmions in the valence (conduction) band
merge. (b)-(c) at $g_m=0.4$ and $g_m=1$, the skyrmions and antiskyrmions separate in the valence and conduction band. For $g_m=1$, by further increasing the disorder strength, the skyrmions generally move towards
band center until on pair of skyrmion and antiskyrmion in the conduction and valence band merges at (d)$W=5$ and then eliminates at (e)$W=7$. (f) In the strong disorder limit, the other pair of skyrmion and antiskyrmion
will merge at $W=11$ and then eliminate. \label{fig4} }
\end{figure}

Next, we provide a phenomenological theory to explain the QAH insulator due to the Berry curvature splitting.
For a nonmagnetic TI with $g_m=0$, the valence bands of the TI carry the Chern numbers $C_{\pm}=\pm1$ due to the nontrivial spin texture, which is analogous to the skyrmions with topological charges $Q_{1,2}^v=\pm1$ in magnetic insulators \cite{QiXL2006}.
The skyrmions with $Q_{1,2}^{v}=\pm1$ ($Q_{1,2}^{c}=\mp1$ ) for valence (conduction) band are schematically plotted as spins pointing outwards and inwards spheres  in Fig~.\ref{fig4}, respectively.
In the absence of magnetization $g_m=0$, one can see two skyrmions of opposite charges in valence (conduction) band
overlap. Then by increasing magnetization strength to $g_m=1$, the two skirmions in valence (conduction) bands with $Q_{1,2}^{v}=\pm1$ ($Q_{1,2}^{c}=\pm1$ ) separate in accordance with the Berry curvature splitting discussed in Fig.~\ref{fig1}. This gives rise to the QAH insulator phase with $C=1$ within two skirmions in the valence (conduction) band, since the total topological charge below the Fermi energy is one.
If we further increase the disorder strength $W$, one pair of skyrmions from conduction and valence bands with opposite topological charges $Q_{1}^{c,v}=\pm1$ will merge and eliminate in pair at the band center $E=0$ [see Fig. \ref{fig4}(d)]. In this circumstance, the two QAH phases within conduction and valence merge together [see Fig.~\ref{fig4}(e)] and the QAH phase survive until the pair of skyrmions merges in Fig.~\ref{fig4}(f).

\begin{figure}[tbh]
	\centering
	\includegraphics[width=3.2in]{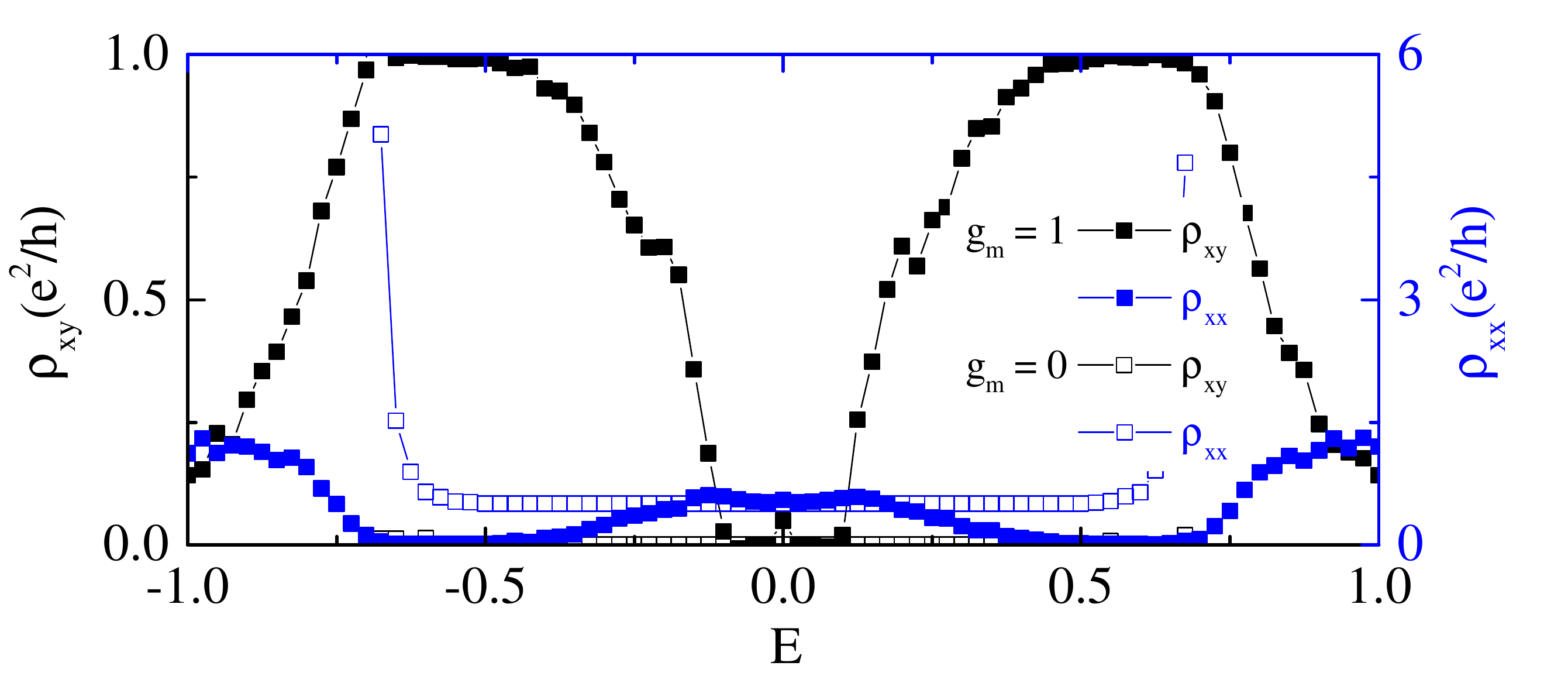}
	\caption{(Color online). The Hall resistance $\rho_{xy}$ (black lines) and longitudinal resistance $\rho_{xy}$ (blue lines) versus the energy $E$ from the standard six-terminal Hall bar measurement. Different symbols represent $g_m=0$ and $g_m=1$. For $g_m=1$, the system shows a reentrant behavior by varying the energy $E$, i.e. from a QAH insulator ($\rho_{xy}=h/e^2$) to a normal insulator ($\rho_{xy}=0$) and then back to QAH insulator ($\rho_{xy}=h/e^2$).
		\label{fig5} }
\end{figure}
{\em Discussion and conclusion.}--
The proposed QAH effect induced by the Berry curvature splitting can, in principle, be experimentally detected by a six-terminal Hall bar. Figure~\ref{fig5} shows the Hall resistance $\rho_{xy}$ and the longitudinal resistance $\rho_{xx}$ versus the energy $E$.
$\rho_{xx}$ and $\rho_{xy}$ are evaluated by the Landauer-B\"{u}tticker formula \cite{datta1995}.
Notably, it is found that the Hall resistance plateaus shows a reentrant behavior for $g_m=1$, namely from ($\rho_{xy}=h/e^2$) to a $\rho_{xy}=0$ and then $\rho_{xy}=h/e^2$, by varying the Fermi energy $E$,
where the longitudinal resistance $\rho_{xx}$ shows two zero-resistance plateaus correspondingly.
This means the system goes through a QAH-normal insulator-QAH transition, which we call the reentrant QAH effect.
The reentrant QAH behavior is distinct from the reentrant quantum Hall effect observed in the TI system \cite{Konig_2008} as well as the QAH effect originating from the band inversion \cite{Yurui2010}.
Therefore, we propose the reentrant QAH effect as a unique feature for the QAH effect
due to the Berry curvature splitting in the presence of moderate disorder.

In reality, due to the localized states in the mobility gap, the quantized Hall conductance  in the QAH can be observed only
when the variable range hopping conductance $g$ is suppressed.
Here  $g(T)\propto\exp[-(T_M/T)^{1/3}]$ \cite{MOTT19681,Tsigankov2002} where $T_M=(B\lambda^2\rho)^{-1}$ with density of states $\rho$, the localization length $\lambda$, the constant $B$ and the temperature $T$.  Thus, the onset temperature of the QAH can be much lower than the inverted gap, since $T_M$ that is usually only a few Kelvin \cite{Feng2015,Ying2016,Deng_2020,yuanbo2020} is much smaller than the energy gap.
To enhance the QAH onset temperature, it is desirable to reduce the density of impurity states $\rho$ and decrease the localization length $\lambda$ of impurity states.

{\emph{Acknowledgement.}}---  We thank Haiwen Liu, Hailong Li and Hua Jiang for illuminating discussions.  This work is financially supported by NBRPC (Grants No. 2015CB921102), NSFC (Grants Nos. 11534001, 11822407, 12074108, and 11704106), and supported
by the Fundamental Research Funds for the Central Universities.
C.-Z. Chen are also funded by the Priority Academic Program Development of Jiangsu Higher Education Institutions. D.-H.X. is also supported by the Chutian Scholars Program in Hubei Province.

%

\end{document}